# Pressure Loss and Sound Generated In a Miniature Pig Airway Tree Model

## Abstract

**Background:** Pulmonary auscultation is a common tool for diagnosing various respiratory diseases. Previous studies have documented many details of pulmonary sounds in humans. However, information on sound generation and pressure loss inside animal airways is scarce. Since the morphology of animal airways can be significantly different from human, the characteristics of pulmonary sounds and pressure loss inside animal airways can be different.

**Objective:** The objective of this study is to investigate the sound and static pressure loss measured at the trachea of a miniature pig airway tree model based on the geometric details extracted from physical measurements.

**Methods:** In the current study, static pressure loss and sound generation measured in the trachea was documented at different flow rates of a miniature pig airway tree.

**Results:** Results showed that the static pressure and the amplitude of the recorded sound at the trachea increased as the flow rate increased. The dominant frequency was found to be around 1840-1870 Hz for flow rates of 0.2-0.55 lit/s.

**Conclusion:** The results suggested that the dominant frequency of the measured sounds remained similar for flow rates from 0.20 to 0.55 lit/s. Further investigation is needed to study sound generation under different inlet flow and pulsatile flow conditions.

**Md Khurshidul Azad\*, Amirtaha Taebi, Joseph H Mansy and Hansen A Mansy**
*University of Central Florida, USA*

**\*Corresponding author:** Md Khurshidul Azad, Biomedical Acoustics Research Laboratory, University of Central Florida, 4000 Central Florida Blvd., Orlando, FL 32816, USA, Tel: +1 (407) 914-1758; E-mail: khurshid@knights.ucf.edu

## Introduction

Several computational and experimental studies of fluid dynamics and acoustic propagation in the pulmonary system were carried out to investigate sound and flow characteristics for various respiratory disease such as Asthma, COPD, and pneumothorax. Since certain pulmonary conditions affect airway diameter and resistance, information about pressure losses in airways may be useful for diagnosis of pulmonary conditions or for patient monitoring. Earlier studies of sound propagation in the airways and lungs demonstrated the utility of these sounds for detecting various pulmonary conditions [1-8]. Sound propagation in the pulmonary system have been studied using animal [3,9,10] and benchtop [11,12] experiments. Numerical models [13-18] were also developed and validated using animal experiments [19-21].

Earlier studies [22,23] measured pressure loss in model of human airways. Olson et al. [24] suggested that the coefficient of resistance due to flow inside airways is dependent on branching angle of a bifurcation; increase in cross sectional area from parent branch to daughter branches, and lengths of each branch in a bifurcation. While pressure and sound inside the airways were widely studied for humans, less information are available for pressure changes and sound generation inside pig airways. It is important to establish information for the airways of certain animal as animal models are often used in medical research. The objective of this research is to study static pressure losses and sound generation of a miniature pig airway tree.

## Materials and Methods

The experimental setup (Figure 1) consists of two axial fans connected in series with a honeycomb between them. The honeycomb helped reduce the turbulence and create axial flow at the second fan inlet. A 100 mm diameter duct was connected to the fan setup outlet and carried the flow into the sound isolation chamber (Model: 4260S, WhisperRoom Inc., Knoxville, TN) where experiments were performed. The duct was then connected to a contraction with inlet and outlet diameter of 100 mm and 10 mm, respectively. The contraction outlet supplied air to an airway model of miniature pig. The airway tree model was built in SolidWorks (2012, SolidWorks corporation, Concord, MA.) based on the geometric features reported in earlier studies of pig airways [7,25-28]. The resulting CAD model was 3D printed using a Desktop 3D printer (MBot3D, El Paso, TX, model: MBot Grid 2) at a print resolution of 0.2mm. A pressure tap was drilled at the trachea 2 cm upstream of the carina to measure the pressure and sound (Figure 2).

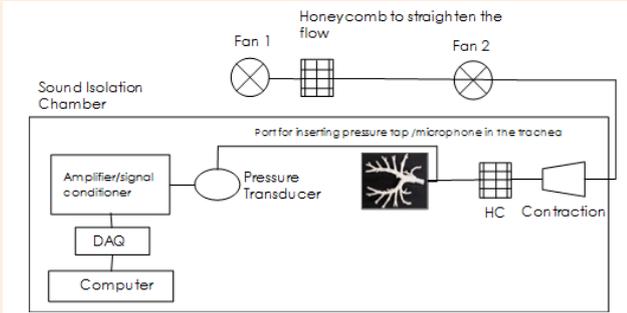

**Figure 1:** Experimental Setup.

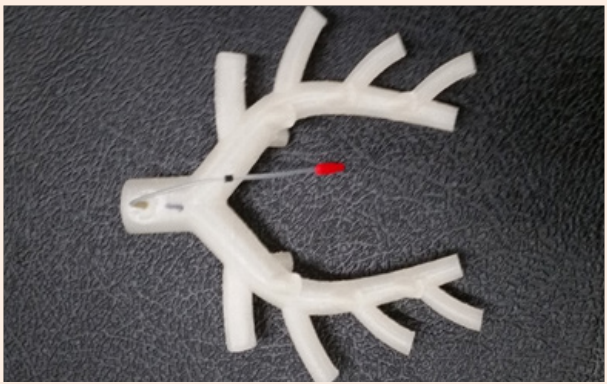

**Figure 2:** Miniature pig airway tree model with small pressure tap at the trachea to measure pressure and sound.

A diaphragm pressure transducer (Model: DP 103, diaphragm: 8-06, Validyne Engineering, CA,) with full scale input of 0.35 cm of water, accuracy of 0.25% of full scale, and sensitivity of about 0.3 V/Pa was used to measure the static pressure at the trachea. The pressure transducer was calibrated using a hot wire anemometer (Model DT-8880, CEM, Shenzhen, China, range: 0.1~25 m/s, resolution: 0.01 m/s, accuracy: ±5% ± 0.1) and a pitot tube (Model: PAA-6-KL, United Sensor Corporation, NH) inside a wind tunnel that has uniform velocity distribution in the test section (except for a thin boundary layer). To calibrate the pressure transducer, the pitot tube was connected to the pressure transducer and placed inside a wind tunnel while the hot wire was placed at the same streamwise location (outside the boundary layer). The measured hot wire velocities at different flow rates were used to calibrate the pressure transducer. Sound was recorded using a probe mic (Model: ER-7C Probe Microphone Series B, Etymotic Research Inc., IL) with a sensitivity: 50mV/Pascal. The static pressure and sound at the trachea were acquired using LabVIEW (Ver 2015, National Instruments, Austin, TX) with a data acquisition card (Model: NI 9215, National Instruments, Austin, TX, resolution: 0.305 mV). The flow rate was measured by placing the airway model inside a thin airbag of maximum volume of 6liters. The time needed to fill the maximum bag volume was recorded using a stopwatch (resolution: 0.01 second). Flow rate was then calculated using the recorded time and airbag volume.

## Results

Hot wire velocities were plotted against the pressure transducer voltages at different flow rates inside the wind tunnel (Figure 3). The velocities were found to be proportional to the transducer output raised to the power of ~0.6, which is close to the theoretical value of 0.5. The velocity values were converted to dynamic pressure using the equation:

$$q = \frac{1}{2}\rho u^2 \quad \dots\dots\dots\dots\dots (1)$$

where $\rho$ and $u$ are the density and velocity of air inside the wind tunnel, respectively.

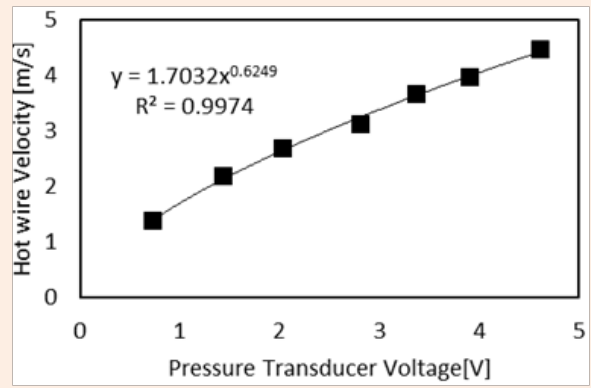

**Figure 3:** Hot wire velocities plotted against pressure transducer voltage. The relation between the velocities and pressure transducer voltages was found to be a power of ~0.6.

Figure 4 shows the relation between the dynamic pressure and pressure transducer voltage. Results showed that the pressure was linearly related to the pressure transducer voltage, which is consistent with the transducer specifications. The Pitot tube was then placed at the exit of the contraction to measure velocity profile at the entrance of the airway tree along two perpendicular axes (in the vertical and horizontal directions).

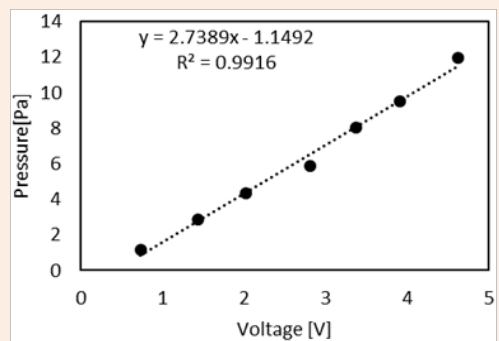

**Figure 4:** Dynamic pressure vs pressure transducer voltage. This relation was used as the calibration equation for the pressure transducer.

Figure 5 shows that the velocity profile was approximately flat at the entrance along both axes. Figure 6 shows the static pressure in the trachea for different flow rates, which equals the pressure loss in the model. The resulting quadratic equation of the regression line shows the relation between pressure loss and flow rate.

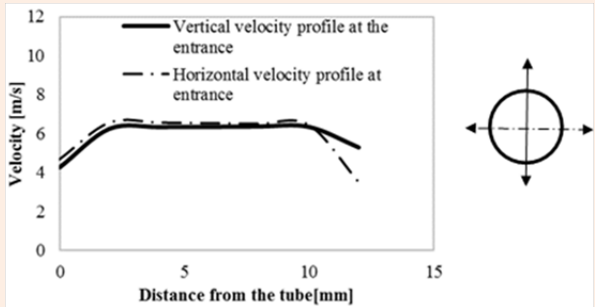

**Figure 5:** Velocity profile measurements at the inlet of the airway tree along the vertical and horizontal axes.

Figure 7 shows the spectrum (using FFT) of the sound recorded at the trachea for different flow rates. Results showed that the dominant frequency ranged from 1840 Hz to1870 Hz for flow rates from 0.20 lit/s to 0.55 lit/s. The sound recorded at no flow condition showed dominant frequency around 120 Hz, suggesting that the background noise level of the system occurs at low frequencies with low amplitudes compared to the sound measured in the model. The spectrum was calculated using FFT. When pulsatile flow is to be studied, time frequency distributions [29,30] may provide a better method to estimate the frequency spectrum.

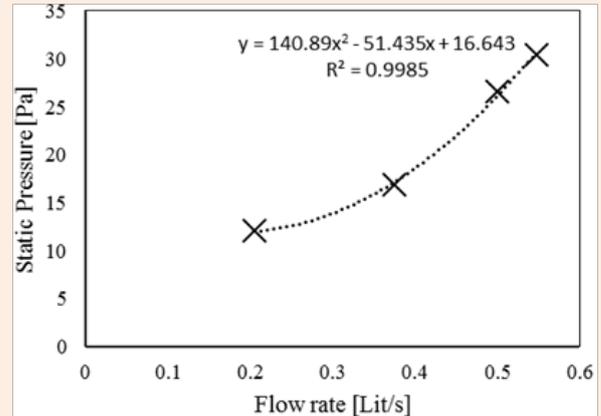

**Figure 6:** Pressure loss in the mode at the trachea for different flow rates.

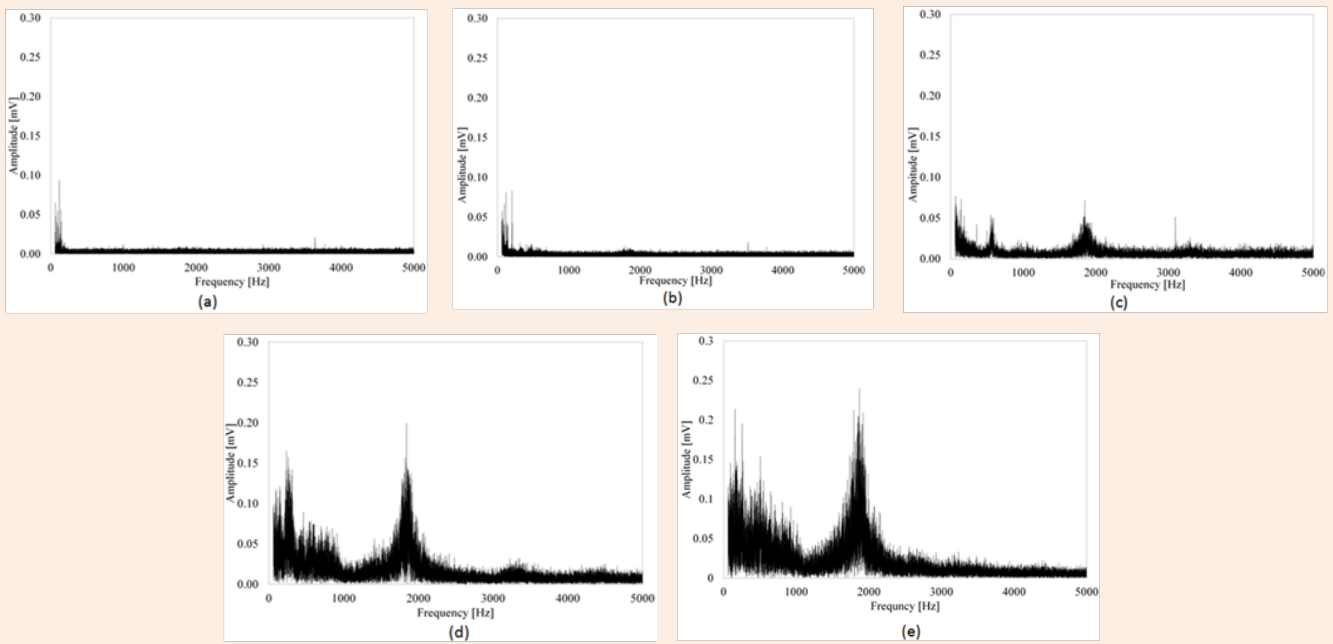

**Figure 7:** Airflow sounds spectral plot at trachea (Figure 2) for flow rate of

a)   0 lit/s (no flow)

b)   0.204 lit/s

c)   0.374 lit/s

d)   0.499 lit/s

e)   0.546 lit/s

## Discussion

The current study used a miniature model of pig airways to investigate the pressure loss and sound generated in the airways. The small diameters of the miniature model may have allowed an increased pressure loss, which can be measured with higher accuracy compared to large diameter models. Results showed that the static pressure at the trachea was small (<30 Pa) for the range of flow rates considered. The pressure difference in individual airway branches is expected to be smaller and a pressure transducer capable of measuring even lower differential pressures would be needed. Hot wire anemometer was chosen as the reference velocity measuring device since it has a calibration certificate provided by manufacturer. The pressure transducer used in this experiment is a diaphragm pressure transducer. The orientation of the pressure transducer can affect the measurements. Hence, the transducer was oriented such that the diaphragm is vertical to help nullify the effect of gravity on the diaphragm. Reducing pressure errors is important especially when measuring low pressure values like those encountered in the current study. The probe microphone was calibrated according to manufacturer's instruction. This was done by connecting the probe microphone to a calibrator supplied by the manufacturer, which generates a 1 kHz 94 dB audio signal. The probe microphone output was measured by the data acquisition system and the signal was found to be within the manufacturer's specifications. In addition, the pressure transducer voltage was found to be linearly proportional to the measured pressure as stated by the manufacturer specifications. When calibrating the pressure transducer, its output was found to be proportional to the square of the velocity, which is consistent with theory. A small pitot tube (diameter= 1.6 mm) was used to minimize disturbance to the flow to be measured.

The velocity profile at the model inlet was predominantly flat along two perpendicular axes. If fully developed inlet flow is desired a longer entrance length would be added. While the flow rate in this experiment was unidirectional and steady, a pulsatile flow will be studied in future investigations to understand the pressure and sound generation characteristics under conditions that are more similar to normal breathing. Figure 7 shows that the amplitude of the sound recorded in the trachea increased as the flow increased but dominant frequencies did not appear to significantly change. The dominant frequency was found to be around 1840~1870 Hz (using FFT) and disappeared when the flow was absent. At no flow condition, the dominant frequency was around 120 Hz, which may be due to the environmental noise or noise in the measurement system. Similar experiments are warranted to study pulsatile flow in airways of different sizes.

## Conclusion

The primary objective of the current study is to investigate the pressure loss and sound generated in a realistic miniature pig airway tree for different flow rates. The results showed that the pressure loss increased with increasing flow rate. The sound measured at the trachea had a dominant frequency around 1840 to 1870 Hz for flow rates of 0.2-0.55 lit/s. Since the airway geometry appears to be significantly different among species, sound generation and pressure loss in pig and human airway trees can be different and, therefore, geometries need to be studied separately. The current study used unidirectional flow with flat inlet velocity profile. Actual tracheal inlet flow is pulsatile and is likely to have complex profiles due to effect of upper airways and vocal cords. Further studies may be warranted to document sound generation under these conditions.

## Acknowledgment

This study was supported by NIH Grant R01 EB012142, R43 HL099053.

## Conflict of Interest

The authors declare no conflict of interest.